\documentclass{aa}  

\usepackage{graphicx}
\usepackage{txfonts}
\usepackage{threeparttable} 
\usepackage{multicol}
\usepackage{natbib,twoopt}
\bibpunct{(}{)}{;}{a}{}{,}             

\begin{document} 

    \title{Alkali lines at extreme densities and their impact on giant planet interior structure}

   \author{L. Siebenaler
          \inst{1}
          ,
          N. F. Allard\inst{2}\fnmsep\inst{3}
          ,
          E. A. van Dijk
          \inst{1}
          \and
          Y. Miguel\inst{1}\fnmsep\inst{4}
          }

   \institute{Leiden Observatory, University of Leiden,
              Einsteinweg 55, 2333CA Leiden, The Netherlands \\
              \email{siebenalerl@strw.leidenuniv.nl}
         \and
             LIRA, Observatoire de Paris, Université PSL, Sorbonne Université, Sorbonne Paris Cité, CNRS, 61 Avenue de l’Observatoire, 75014
            Paris, France
         \and
            Institut d’Astrophysique de Paris, UMR7095, CNRS, Université Paris VI, 98bis Boulevard Arago, 75014 Paris, France
         \and
             SRON Netherlands Research Organisation Netherlands, Niels Bohrweg 4, 2333CA Leiden, The Netherlands 
             }

   \date{Received 26 June 2026; accepted 23 July 2026}

  \abstract
   {Alkali lines, in particular the sodium Na $D$ (5891 $\AA$, 5897 $\AA$) and potassium K $D$ (7667 $\AA$, 7701 $\AA$) resonance doublets, are dominant opacity sources in giant planets over a wide range of temperatures ($\gtrsim 1000  \ \rm K$). Their  strong pressure-broadened wings significantly influence the thermal structure of giant planets, especially at high pressures. Most detailed line-profile calculations have so far been limited to perturber densities up to $10^{21} \ \rm cm^{-3}$. However, conditions in the deep atmospheres and interiors of giant planets can reach significantly higher densities, making the temperature gradients increasingly uncertain.}
   {We determined how physically consistent collisional broadening of the  Na $D$ and K $D$ lines at extreme densities affects opacity calculations and consequently the inferred interior structure of giant planets.}
   {We computed detailed Na $D$ and K $D$ line profiles using unified line theory, extending to molecular hydrogen perturber densities of $n_{\rm H_2} = 5 \times 10^{22} \ \rm cm^{-3}$, which translates to pressures up to $\sim 20 \ \rm kbar$. The revised cross sections were incorporated into Rosseland mean opacity tables, which were then used to evaluate their effect on planetary thermal structures.}
   {At densities $n_{\rm H_2} > 10^{21} \ \rm cm^{-3}$, the line profiles predicted by unified line theory exhibit significantly stronger and more extended wings than commonly used Voigt profiles, as well as density-dependent line shifts. The revised line profiles increase Rosseland mean opacities by a factor of 2 at $10^3 \ \rm bar$ and by an order of magnitude at $10^4 \ \rm bar$. Consequently, the radiative-convective boundary of warm and hot giant planets can shift to lower pressures, producing warmer interior adiabats and increasing inferred core masses by up to $9 \ \rm M_{\rm earth}$. We further find that Jupiter is highly unlikely to host a stable radiative layer at the present time or throughout most of its evolution, as the required Na and K abundances for this are well below observational constraints. Our new opacity tables span metal mass fractions $Z \in [0.0044, 0.696]$ and are publicly available.}
   {}

\titlerunning{Alkali lines at extreme densities and their impact on giant planet interior structure}
\authorrunning{Siebenaler et al.}
\maketitle
%

\section{Introduction}
The absorption lines of the sodium (Na) $3s-3p$ and potassium (K) $4s-4p$ resonance doublets are prominent features in the optical spectra of late-type stars and brown dwarfs (e.g., \citealt{Burrows_2000, Rodrigo_2007}), as well as giant planets (e.g., \citealt{Charbonneau_2002, Snellen_2008, Sing_2016}). This is due to their extremely strong pressure-broadened wings arising from collisions with molecular hydrogen (H$_2$) and helium (He), which can act as a pseudo-continuum opacity and in the case of K extend into the near-infrared. Extensive studies of these alkali line profiles have been carried out over the past decades (e.g., \citealt{allard1982, Burrows_2002, Burrows_2003, Allard_2016, Allard_2019}), enabling their implementation in atmospheric radiative transfer models up to perturber densities of $10^{21} \ \rm cm^{-3}$.\footnote{For temperatures between $1000 - 3000 \ \rm K$, perturber densities of $10^{21} \ \rm cm^{-3}$ correspond to pressures of approximately $\sim 100 - 400 \ \rm bar$.} The accurate treatment of these line profiles is essential, as they can strongly influence the atmospheric energy balance, potentially leading to temperature inversions \citep{Molliere_2015} and affecting the thermal structure deeper in the atmosphere \citep{Baudino_2017}. The reason is that in H$_2$-dominated atmospheres, there exists a window between temperatures of $\sim 1400 - 2100 \ \rm K$, where in the absence of disequilibrium chemistry these lines provide the dominant gaseous opacity sources at short wavelengths ($\lambda \lesssim 1 \ \mu \rm m$), while other short-wavelength absorbing species, such as metal hydrides, metal oxides, and free electrons, become abundant only at higher temperatures.

Even for the interior of giant planets, these lines play an important role, as they can dictate the dominant heat transport mechanism. In \cite{Siebenaler_2025} (hereafter S25), we showed that a sufficiently strong depletion in Na and K can lead to the formation of a stable radiative layer in Jupiter around the kilobar level (see also \citealt{Guillot_1994, Guillot_2004}). The presence of such a layer has been discussed in the context of Juno observations \citep{Bolton_2017}, as it could suppress vertical mixing and help reconcile Jupiter’s low atmospheric CO abundance with its inferred high deep water abundance \citep{Cavalie_2023}. It may also contribute to resolving the so-called Jupiter $Z$ problem, i.e., the tension between atmospheric metallicity constraints and interior structure models \citep{Howard_2023b, Muller_2024, Nettelmann_2025}.

Despite their importance, modeling Na $D$ and K $D$ line profiles at high pressures remains challenging. Most detailed studies on their line profiles are limited to perturber densities up to $ 10^{21} \ \rm cm^{-3}$, which is sufficient for interpreting the presence of alkali in observed spectra, but not for modeling deep atmospheres and interiors. In S25, line profiles at higher densities were approximated using the impact theory \citep{Lorentz_1906, Weisskopf_1932}, which treats collisions as instantaneous and leads to Lorentzian line shapes. However, this approach fails not far from the line center and as a result inaccurately describes line wings. A common approach to mitigate these issues is to introduce ad hoc line-wing cutoffs. However, this is not physically well justified, and line wings are expected to grow in strength with increasing pressure and therefore cannot be ignored indefinitely.

An alternative description is provided by the quasi-static theory \citep{Kuhn_1934, Kuhn_1937}, which no longer treats collisions as instantaneous, but instead accounts for the interaction potential between the absorbing atom and perturber during collisions, while assuming that the perturber is effectively static. This approach provides a more accurate description of line wings, but it fails to reproduce the line cores, which are produced by both distant and close encounters.

Unified line theory provides a more consistent framework by simultaneously describing both the line cores and wings (e.g., \citealt{anderson1952, allard1978, royer1978, allard1999}). While it has been successfully applied to the Na $D$ and K $D$ lines at moderate densities, its application to higher densities relevant to deep atmospheres and interiors has remained limited. \cite{Allard_2025} recently applied unified line theory to the Na and K $P_{3/2}$ ($D2$) lines up to H$_2$ perturber densities of $2 \times 10^{22} \ \rm cm^{-3}$ at a temperature of 1000 K, demonstrating that the impact theory can strongly underestimate line widths and fails to capture density-dependent line shifts.\footnote{$P_{3/2}$ refers to the quantum state involved in the transition. Here, the electron has an orbital angular quantum number $L = 1$, and a total angular momentum quantum number $J = 3/2$.}

In this work, we extend the calculations of \cite{Allard_2025} to both the $P_{1/2}$ ($D1$) and $P_{3/2}$ ($D2$) Na and K lines over a range of temperatures $1000 - 3000 \ \rm K$ and up to H$_2$ perturber densities of $5 \times 10^{22}\ \rm cm^{-3}$, which corresponds to pressures $\sim 7 - 20  \ \rm kbar$. These calculations enable improved opacity treatments for the interiors and deep atmospheres of giant planets. We include the revised Na $D$ and K $D$ cross sections in our opacity tables from \cite{Siebenaler_2026} (hereafter SM26) and explore their impact on the inferred interior structure of giant planets. In Sect. \ref{sec:Methods}, we briefly describe unified line theory and the calculations of Rosseland mean opacities. In Sect. \ref{sec:Results}, we present the revised line profiles and compare them to predictions from the impact theory, as well as their effect on mean opacities. In Sect. \ref{sec:Discussion}, we discuss the implications for planetary thermal structures, including the location of the radiative–convective boundary (RCB) in warm and hot Jupiters, and we revisit the conditions for a stable radiative layer in Jupiter. Finally, in Sect. \ref{sec:Conclusion}, we give our conclusions.

\section{Methods} \label{sec:Methods}

\subsection{Unified line theory}

In this work, we compute detailed Na $D$ and K $D$ line profiles at extreme densities and study their effect on opacity calculations. To this end, we use unified line theory treatment, which accounts for the finite duration of collisions and enables to model line profiles from their centers to their far wings. This approach overcomes the key limitation of the impact theory, which assumes that collisions occur instantaneously and therefore breaks down not far away from the line center. We first introduce the general expression for the spectrum in unified theory, together with the main quantities entering the method. This is followed by a brief description of the formation of line satellites, a prominent feature of the Na $D$ and K $D$ lines.

\subsubsection{General expression for the spectrum in an adiabatic representation} \label{sec:methods_autocorrelation_function} 
In unified theory, the shape of pressure-broadened alkali absorption lines from near resonance to the far wing is obtained using an autocorrelation formalism. The fundamental result expressing the autocorrelation function for many perturbers in terms of a single perturber quantity $g(s)$ was first obtained by \citet{anderson1952} and \citet{baranger1958a}  in the classical and quantum cases respectively. In this formalism, we write the spectrum $I(\Delta\omega)$ as the Fourier transform (FT) of the dipole autocorrelation function $\Phi(s)$,

\begin{equation}
I(\Delta\omega)=
\frac{1}{\pi} \, Re \, \int^{+\infty}_0\Phi(s)e^{-i \Delta\omega s} ds,
\label{eq:int}
\end{equation}
where $s$ is time.
The FT in Eq.~(\ref{eq:int}) 
is taken such that $I(\Delta\omega)$ is
normalized to unity when integrated over all frequencies,
and $\Delta\omega$ is measured relative to  the 
unperturbed line.
The dipole autocorrelation function $\Phi (s)$ is evaluated for
a classical collision path with an average over all possible collisions.
Complete details and the derivation of the theory 
are given by \citet{allard1999}.
The well-developed theory of spectral line shapes allows us to compute
the function
\begin{equation}
\Phi(s) = e^{-n_{\rm p}g(s)}
\label{eq:phi}
\end{equation}
in which the density of perturbers $n_{\rm p}$ is expressed explicitly,
and the function $g(s)$ depends only on single collisions.
The decay of the autocorrelation function $\Phi (s)$ with time leads to
atomic line broadening. It depends on the density of perturbing atoms or molecules
$n_p$ and on their interaction with the radiating atom.
In radiative collision transitions, it is the
difference potential $\Delta V(R)$ between the
final and initial states that determines the frequency and the energy emitted or absorbed by a single photon. The potentials and radiative dipole transition moments are input data which
are now known with high accuracy when using ab initio potentials.

\begin{figure}[!h]
    \centering
    \includegraphics[width=1\linewidth]{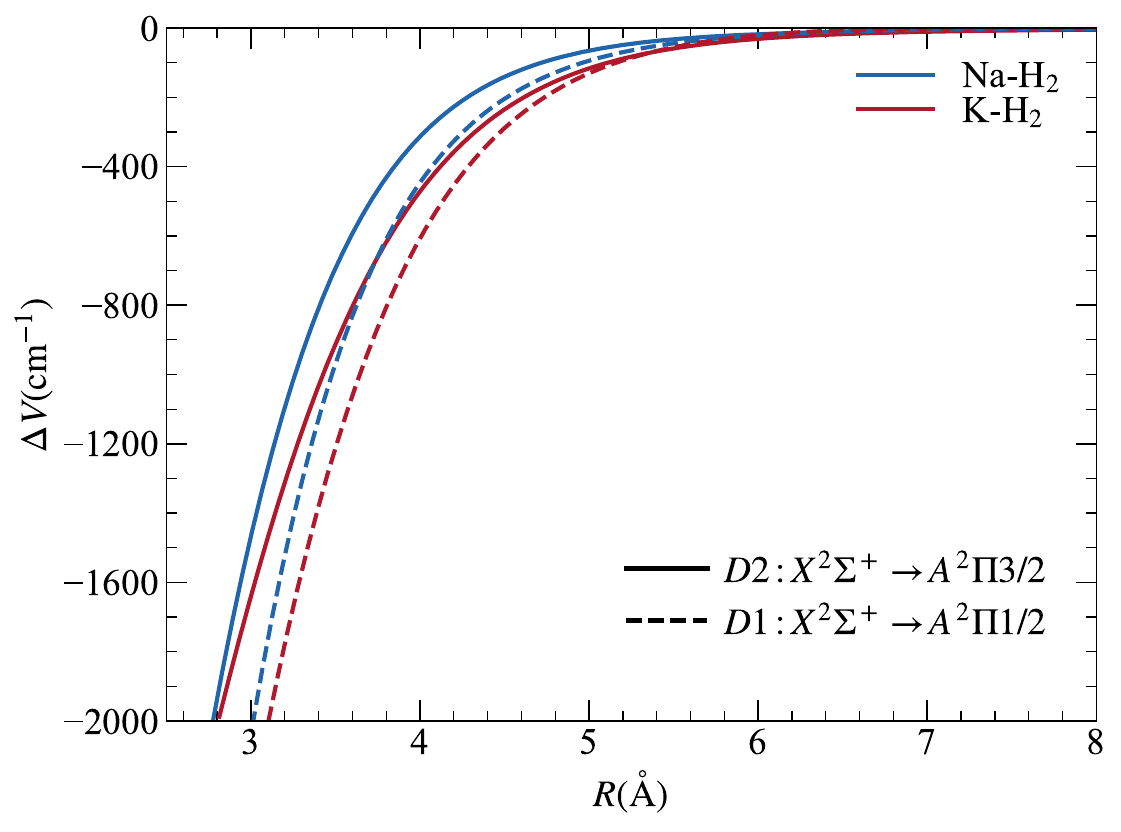}
    \caption{$\Delta V(R)$  for the transitions
  \hbox {$ X ^2\Sigma^+ \rightarrow A ^2\Pi1/2$} (dashed curves),
   \hbox {$ X ^2\Sigma^+ \rightarrow A ^2\Pi3/2 $} (full curves),
   involved in the formation of the red wings
   of  Na-H$_2$ (blue curves), K-H$_2$ (red curves).}
    \label{fig:DeltaV_red}
\end{figure}

\begin{figure}[!h]
    \centering
    \includegraphics[width=1\linewidth]{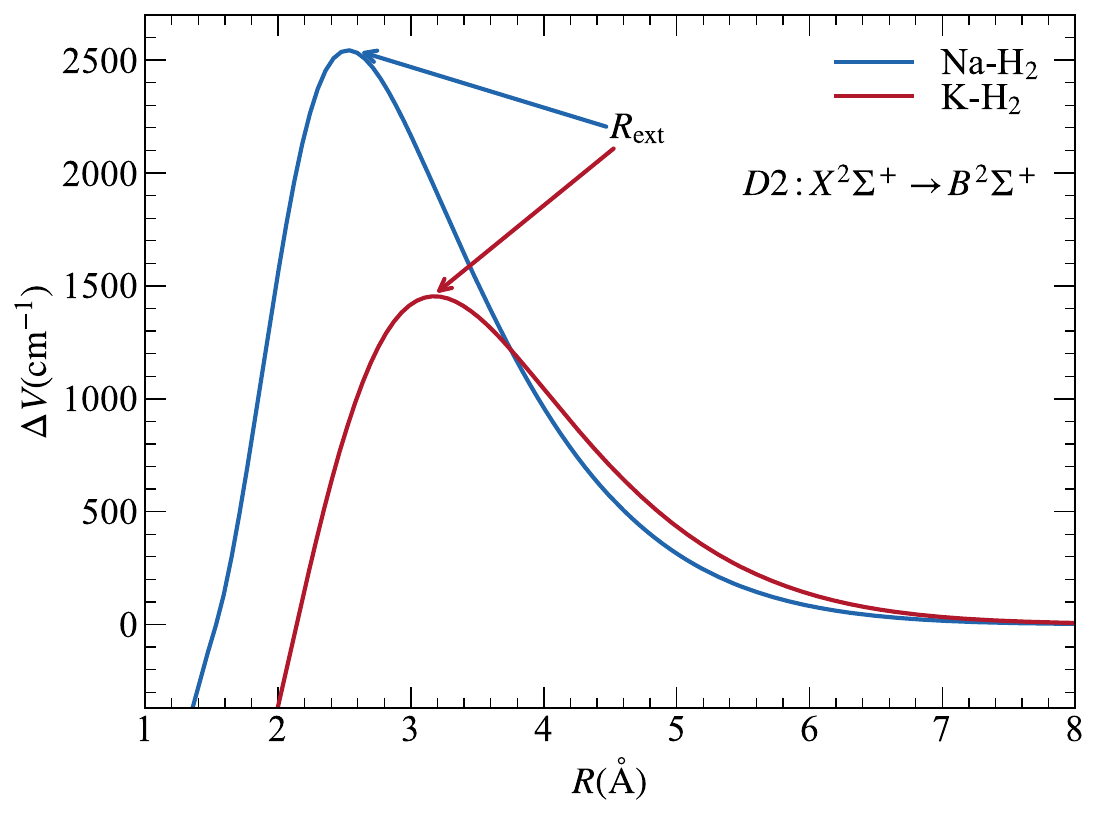}
    \caption{$\Delta V(R)$  for the transitions
  \hbox {$ X ^2\Sigma^+ \rightarrow B ^2\Sigma^+$}
  involved in the formation of the Na-H$_2$  line satellites (blue curve)
  and K-H$_2$ line  satellites (red curve).}
    \label{fig:DeltaV_blue}
\end{figure}

In this work,  we  use the data of \citet{Allard_2019} for the resonance lines of Na-H$_2$ and of \citet{Allard_2016} for K-H$_2$ to compute their autocorrelation functions $\Phi(s)$ at temperatures $T =$ 1000, 2000 and 3000 K, thus covering the range where they are expected to be the most important. We compute the FT of the respective $\Phi(s)$ for H$_2$ perturber densities ranging from $n_{\rm H_2} = 10^{21}  - 5 \times 10^{22} \ \rm cm^{-3}$. We show the adopted $\Delta V(R)$ for the resonance lines in Figs.~\ref{fig:DeltaV_red}-\ref{fig:DeltaV_blue}. For simplicity, we only show $\Delta V$ for the symmetry $C_{2v}$. For other symmetries $\Delta V$ exhibits similar behaviors and is of the same order (see Fig. 8 in \citealt{allard2007}). The $P_{1/2}$ ($D1$) line is due to a simple isolated $A$ $\Pi_{1/2}$ state, whereas the $P_{3/2}$ ($D2$) line comes from the $A$ $\Pi_{3/2}$ and $B$ $\Sigma_{1/2}$ adiabatic states arising from the $4p$ $P_{3/2}$ atomic state.  While $B$ states radiate in the blue wing, $A$ states radiate in the red wing. 
The radiation of $A$ states in the red wing can be directly inferred from the difference potential $\Delta V$ in Fig. \ref{fig:DeltaV_red}, which is negative over all Na-H$_2$ and K-H$_2$ separations $R$. The radiation of the $B$ states in the blue wing is associated with the formation of satellite bands, which will be introduced in the following part.

\begin{figure}[!h]
    \centering
    \includegraphics[width=1\linewidth]{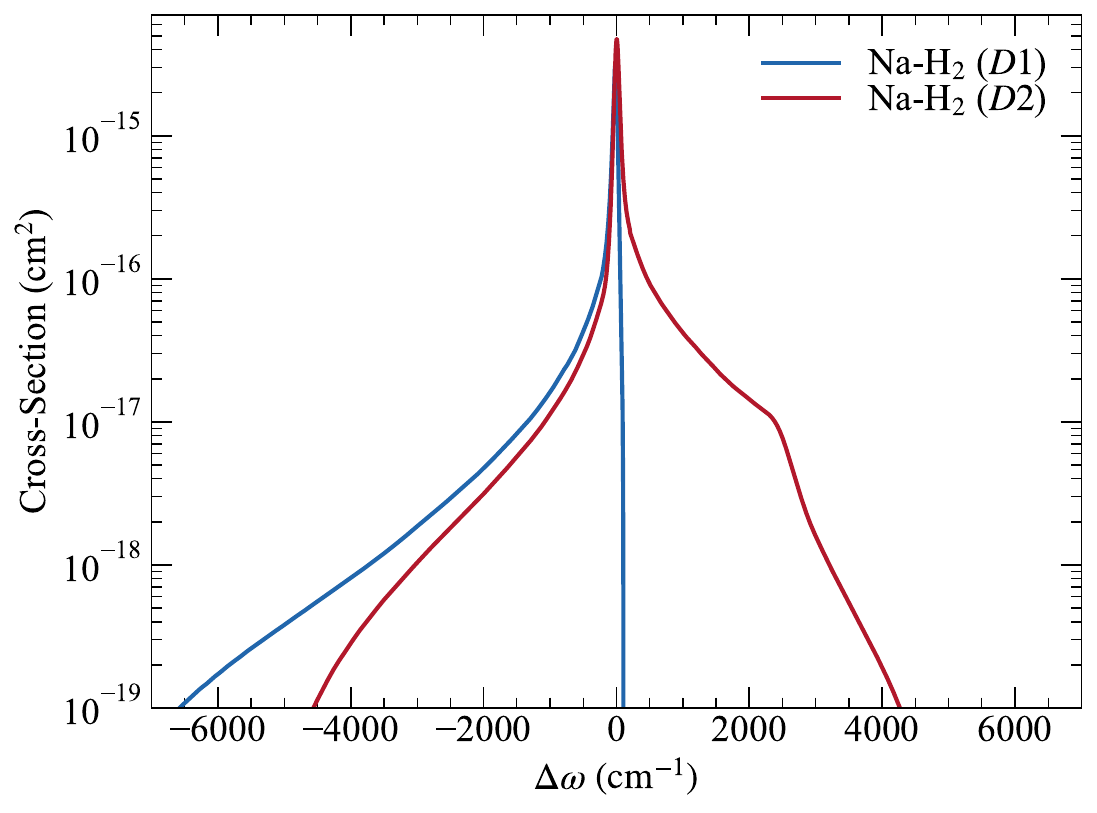}
    \caption{Line profiles of the Na-H$_2$ $D1$ (blue) and $D2$ (red) lines at a perturber density $n_{\rm H_2} = 10^{21} \ \rm cm^{-3}$ and temperature $T=2000 \ \rm K$.}
    \label{fig:Na-H2_1e21}
\end{figure}

\subsubsection{Formation of line satellites} \label{sec:satellites}
The line wing, generally, does not decrease  monotonically with increasing
frequency separation from the line center. Close collisions between a
radiating  atom and a perturber are responsible for transient quasi-molecules 
which may lead to the appearance of satellite features in the  wing of
an atomic line profile. Its shape, for an atom in the
presence of other atoms is sensitive to the difference  between the initial
and final state interaction potentials $\Delta V(R)$. When $\Delta V (R)$, for a given
transition, goes through an extremum, a relatively wider range of interatomic
distances contribute to the same spectral frequency, resulting in an
enhancement, or  satellite, in the line wing.
These structures in the wings of atomic lines can be used as
diagnostics of the stellar temperature and gravity. 
Their characteristics (position, amplitude and shape), due to the
formation of quasi-molecules during collisions between
the radiating atom and perturbers, depend directly on the potential
energy curves correlated to the atomic levels of the transition.
Unified theory \citep{anderson1952,allard1978,royer1978}
predicts that there will be line satellites  centered periodically at 
frequencies corresponding to the extrema of the difference potential
between the upper and lower states, 

\begin{equation}
 \Delta\omega=~k_{\rm p}\Delta V_{{\rm ext}},
\label{eq:sat}
\end{equation}
 ($k_{\rm p}$=1,2,3,\ldots).
Here $\Delta\omega$ is the frequency difference between the center of
the unperturbed spectral line and the satellite feature.
This series of line  satellites corresponds to the simultaneous presence of $k_{\rm p}$ perturbers in the collision volume $\mathcal{V}_{\mathrm{coll}}$. Multiple line satellites at very high densities were first identified in the pioneering work of \cite{mccartan1969}, with a definitive experimental observation later reported by \cite{kielkopf1979}.

Blue satellite bands in the Na-H$_2$ and K-H$_2$ $D2$ profiles can be predicted from the
extrema in the  potential difference $\Delta V$  related to the 
\hbox {$ X ^2\Sigma^+ \rightarrow B ^2\Sigma^+$}
transition. 
$\Delta V$ reported in Fig. \ref{fig:DeltaV_blue} have  a maximum
$\Delta V_{\rm ext}=2500$~cm$^{-1}$ at $R=2.5$ ~\AA\/ for Na-H$_2$ and
$\Delta V_{\rm ext}=1450$~cm$^{-1}$ at $R=3.2$~\AA\/ for K-H$_2$. As a result, the $D2$ lines show enhanced absorption in the blue wings. In addition, the $D2$ lines exhibit strong red wings due to the contribution from the $A$ state, although these do not give rise to satellite features. In contrast, the $D1$ lines only originate from an isolated $A$ state and therefore do not exhibit satellite bands. They are strongly asymmetric with a strong red wing and weak blue wing. An example of these features in the Na-H$_2$ $D1$ and $D2$ line profiles is shown in Fig. \ref{fig:Na-H2_1e21}.

\begin{figure*}[!h]
    \centering
    \includegraphics[width=0.9\linewidth]{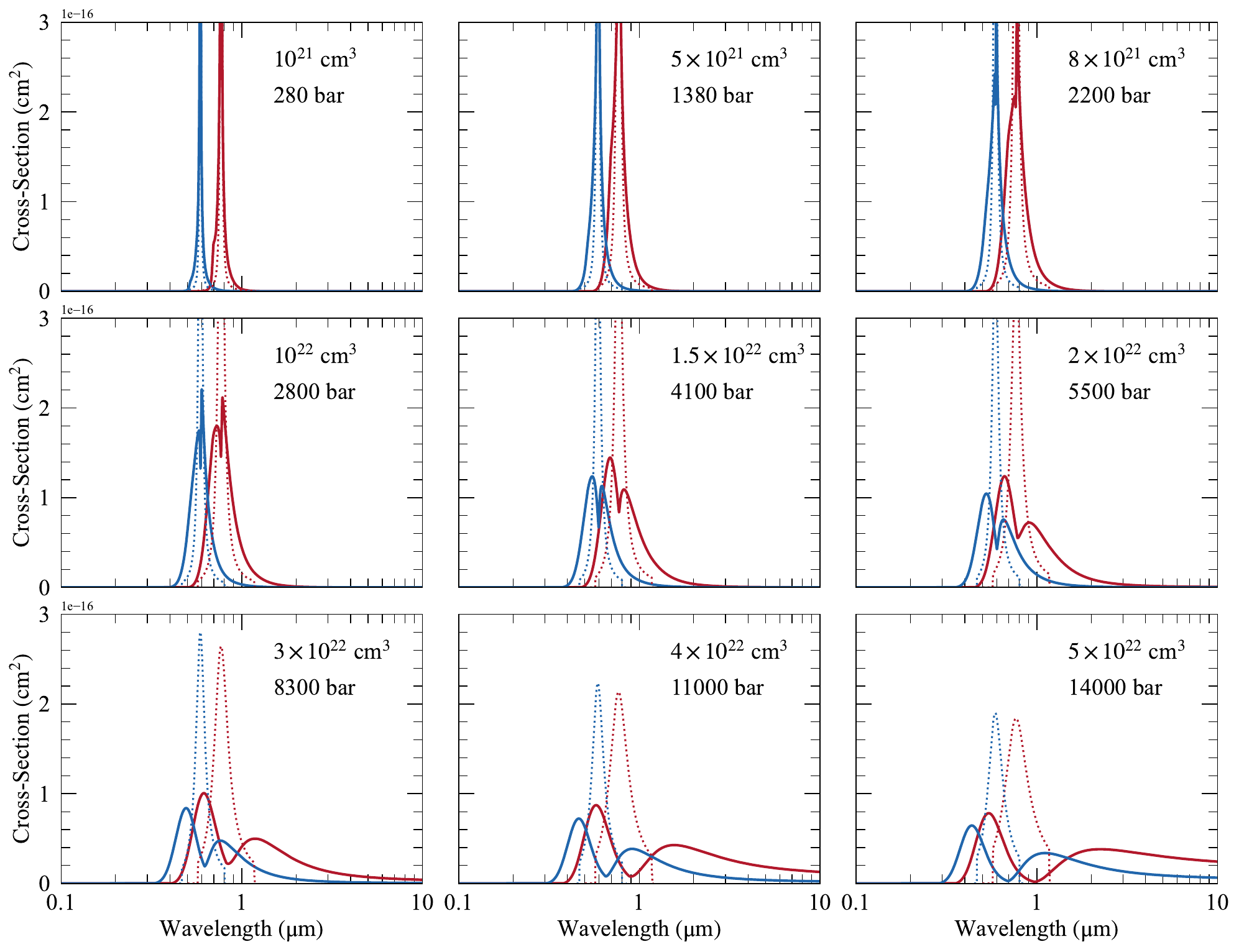}
    \caption{Cross sections of the Na $D$ (blue) and K $D$ (red) line profiles at different H$_2$ perturber densities $n_{\rm H_2}$, for a fixed temperature of $T = 2000 \ \rm K$. The solid curves show profiles computed using unified line theory, while dotted curves represent Voigt profiles with a line wing cutoff of $4500 \ \rm cm^{-1}$. }
    \label{fig:profiles_NaD_KD}
\end{figure*}

\subsection{Rosseland mean opacity}
In this study, we focus on the impact of our revised Na $D$ and K $D$ line profiles on the interior structure of giant planets. We quantify this using the Rosseland mean opacity $\kappa_{\rm R}$, which is the relevant parameter to describe the temperature gradients in the interior of planets or stars. In contrast, the Planck mean opacity, $\kappa_{\rm P}$, is more appropriate for optically thin media. We therefore exclusively focus on $\kappa_{\rm R}$ throughout this study. We next show how $\kappa_{\rm R}$ enters the description of planetary interiors.

To determine the thermal structure of planets, the radiative transfer equation must be solved. In planetary atmospheres this problem is computationally expensive and typically requires numerical models due to the frequency dependent nature of opacities. However, in the interior or deep atmosphere of planets, the medium becomes optically thick and radiative transfer can be significantly simplified. A large body of literature discusses this limit in detail, but here we only want to briefly outline the diffusion approximation and demonstrate how the $\kappa_{\rm R}$ enters the solution (e.g., \citealt{Rybicki_1986}). The diffusion approximation applies when the radiation field is nearly isotropic and local thermodynamic equilibrium holds. Under these conditions, the total energy flux through a layer at radius $r$ can be written as

\begin{equation} \label{eq:tot_energy_flow}
    F(r) = -\frac{4\pi}{3\rho}\int_{0}^{\infty} \frac{1}{\kappa_{\nu}} \frac{\textrm{d}B_{\nu}}{\textrm{d}r} \textrm{d}\nu, 
\end{equation}

where $\rho$ is the mass density, $\kappa_{\nu}$ is the frequency dependent opacity, and $B_{\nu}$ is the Planck function. Rewriting Eq. (\ref{eq:tot_energy_flow}) in terms of the temperature gradient $\textrm{d}T/\textrm{d}r$ and using $\int_0^{\infty} \textrm{d}B_\nu/\textrm{d}T  \textrm{d}\nu = 4\sigma_{\rm SB} T^3/\pi$ gives

\begin{equation} \label{eq:tot_energy_flux_2}
    F(r) = -\frac{16 \sigma_{\rm SB}  T^3}{3\rho \kappa_{\rm R}} \frac{\textrm{d}T}{\textrm{d}r},
\end{equation}

where $\sigma_{\rm SB}$ is the Stefan-Boltzmann constant and

\begin{equation}
    \frac{1}{\kappa_{\rm R}} = \frac{\int_0^\infty \frac{1}{\kappa_\nu} \frac{\textrm{d} B_\nu}{\textrm{d}T} \textrm{d}\nu}{\int_0^\infty \frac{\textrm{d} B_\nu}{\textrm{d}T} \textrm{d}\nu}
\end{equation}

is the inverse of the Rosseland mean opacity. It corresponds to a harmonic mean of the monochromatic opacity $\kappa_\nu$, and consequently the energy transport is controlled primarily by the most transparent parts of the opacity spectrum. Equation (\ref{eq:tot_energy_flux_2}) shows that in optically thick regions the thermal structure can be obtained using only the mean opacity $\kappa_{\rm R}$, avoiding the need to solve the radiative transfer equation for each wavelength. This corresponds to the approach for how evolution models of stars (e.g., \citealt{Paxton_2011, Manchon_2025}) or planets (e.g., \citealt{Guillot_1994, Sur_2024}) determine the thermal structure as they rely on precalculated $\kappa_{\rm R}$ tables (e.g., \citealt{Freedman_2014}; SM26). In this study, we exclusively consider high densities regions, where it is safe to assume that the diffusion approximation holds. Hence, unless stated otherwise, we use $\kappa_{\rm R}$ to determine the thermal structure of planets.

Given that $\kappa_{\rm R}$ represents an averaged opacity, it may appear surprising that the detailed shape of the Na $D$ and K $D$ resonance lines can play an important role. As shown by \cite{Guillot_2004} and S25, for Jupiter in the temperature range $\sim 1400 - 2100 \ \rm K$, $\kappa_{\rm R}$ is largely controlled by absorption associated with these alkali lines. 
In the absence of these lines a broad transparent window appears in the opacity spectrum at optical wavelengths which will strongly influence $\kappa_{\rm R}$. It is the broad line wings of the Na $D$ and K $D$ that can effectively fill this opacity window. However, as shown in S25, differences in the extent of these line wings can lead to substantial changes in $\kappa_{\rm R}$ and consequently alter the calculated thermal structure of the planet. This suggests that an accurate description of the alkali line profiles is required even when considering $\kappa_{\rm R}$.

\begin{figure*}[!h]
    \centering
    \includegraphics[width=0.9\linewidth]{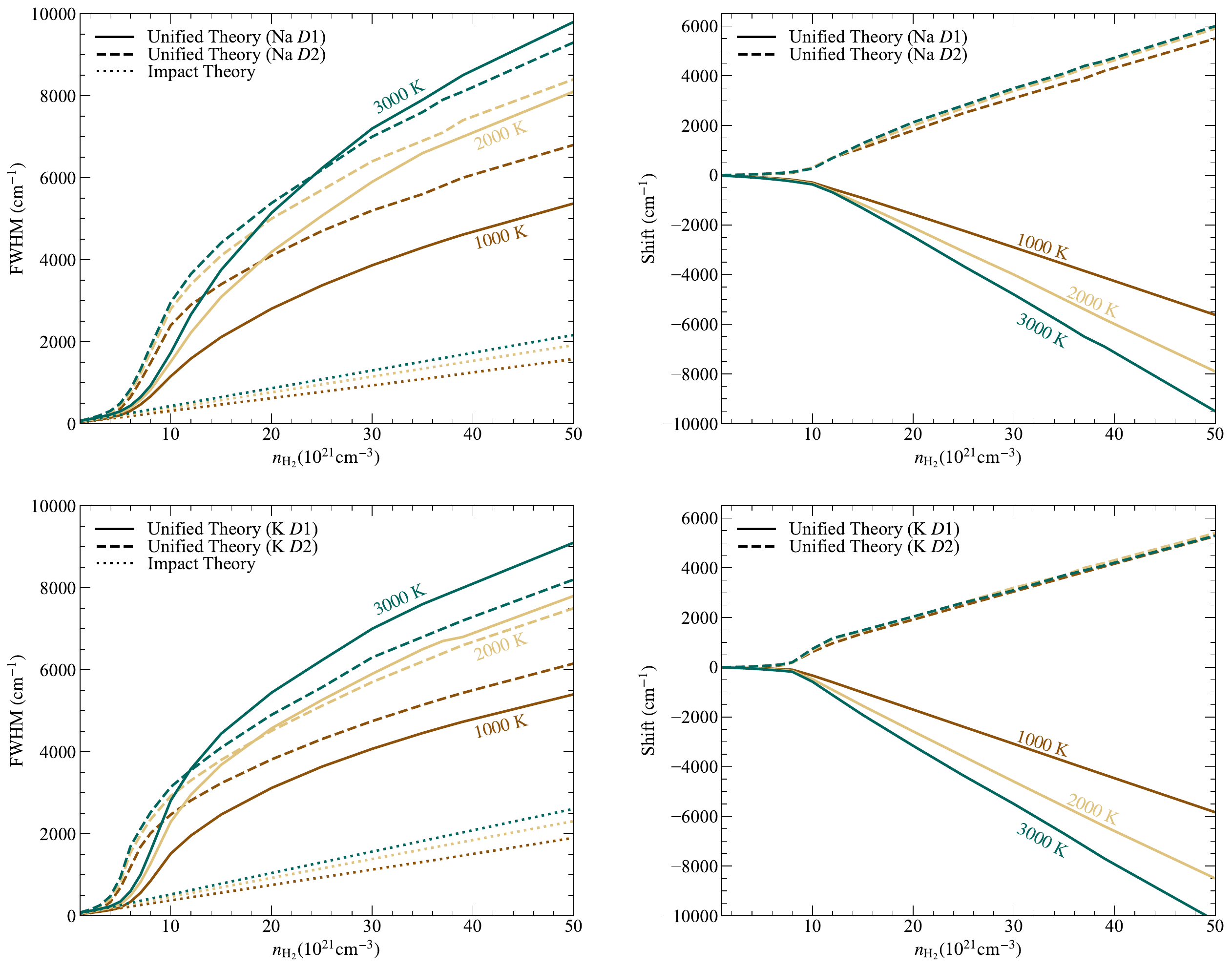}
    \caption{Full width at half maximum (FWHM; left panels) and shift (right panels) of the Na $D$ (top) and K $D$ (bottom) line profiles as a function of H$_2$ perturber density $n_{\rm H_2}$. Colors indicate the temperature. The solid and dashed curves correspond to the prediction from unified line theory of the $D1$ and $D2$ lines respectively. The dotted curves give the FWHM prediction from the impact theory.}
    \label{fig:FWHM_Shift_Nad_KD}
\end{figure*}

\subsection{Mean opacity table}

In this work, we are incorporating our new high-density Na $D$ and K $D$ absorption cross sections in the SM26 mean opacity tables.\footnote{We update both $\kappa_{\rm R}$ and $\kappa_{\rm P}$ in our tables, although the conditions of our revised line profiles are in principle not relevant for the application of $\kappa_{\rm P}$.} These tables already rely on detailed Na $D$ and K $D$ line profiles derived from unified line theory (\citealt{Allard_2016, Allard_2019, Allard_2025}) which extend to perturber densities of $n_{\rm H_2}= 10^{21} \ \rm cm^{-3}$. At higher densities, the line profiles are modeled by a Voigt profile with a line-wing cutoff of $4500 \ \rm cm^{-1}$. In the present work, we replace this high-density treatment with our new calculations. A summary of all the gaseous cross section sources considered in this work for the mean opacity tables is given in Appendix \ref{Appendix:xsec}.

As mentioned in Sect. \ref{sec:methods_autocorrelation_function}, we have computed $\Phi(s)$ for the Na $D1$ and $D2$, and K $D1$ and $D2$ lines perturbed by H$_2$ at temperatures $T=1000$,  $2000$, and $3000 \ \rm K$. To obtain cross sections at the temperature and pressure grid-points of the SM26 tables, we perform bilinear interpolation using linear spacing in temperature and logarithmic spacing in perturber density. We do not perform extrapolation. For temperatures above 3000 K (below 1000 K), we interpolate only in density while adopting $\Phi(s)$ computed at $T = 3000 \ \rm K$ ($T = 1000 \ \rm K$). Similarly, for perturber densities above $n_{\rm H_2} > 5\times 10^{22} \ \rm cm^{-3}$ we interpolate only in temperature space. To translate the H$_2$ perturber density to pressure, we assume that the H$_2$ partial pressure is equal to the total atmospheric pressure, i.e. H$_2$ is the only broadening species. In doing so, we slightly overestimate the broadening due to H$_2$ and neglect any additional broadening from He. However, since giant planet atmospheres are strongly dominated by H$_2$, this approximation is expected to have only a minor effect.

While the treatment above updates the Na and K absorption cross sections, the corresponding neutral alkali abundances used in constructing our opacity tables are obtained from a chemical-equilibrium calculation that does not include nonideal effects. We use the equilibrium chemistry code \texttt{GGchem} \citep{Woitke_2018}, in which the law of mass action is formulated for an ideal-gas mixture. At the high densities considered here, nonideal effects could alter the 
neutral Na and K number densities and level populations entering our opacity calculations. Although the magnitude of
these effects is difficult to quantify, the analysis of
\cite{Marigo_2024} suggests that ionization-potential depression and
pressure ionization, in particular, are unlikely to have a significant effect due to the relatively low temperatures considered here. A self-consistent treatment of nonideal chemistry is beyond the scope of the present work and should be addressed in future studies.

\section{Results} \label{sec:Results}

In this section, we present the Na $D$ and K $D$ line profiles as predicted by unified line theory, and show how they change $\kappa_{\rm R}$. Our new mean opacity tables have the same format as those in SM26 and are available in the updated Zenodo repository. They span metal mass fractions $Z \in [0.0044, 0.696]$, which translates to metal abundances $\rm [M/H] \in [-0.5, +2.21]$. 

\subsection{Line profiles: Unified theory versus impact theory}

We now present the Na $D$ and K $D$ line profiles computed using unified line theory and compare them to profiles based on the impact theory. Specifically, we compare our results to the cross section data used in S25 and SM26, where the Na $D$ and K $D$ lines are modeled as Voigt profiles with a line wing cutoff of 4500 $\rm cm^{-1}$ for perturber densities $n_{\rm H_2} \geq 10^{21} \ \rm cm^{-3}$. One of the main differences introduced by unified line theory is that the resulting profiles are intrinsically asymmetric. This arises from the difference potentials $\Delta V$, which are a key input to unified theory and were determined from ab initio calculations. As explained in Section \ref{sec:satellites}, they cause the $D2$ lines to exhibit both strong red and blue wings, with the blue wings being characterized by satellite bands, while the $D1$ lines have only strong red wings. In contrast, the impact theory always approximates lines as Lorentzian and cannot capture any asymmetries.

Figure \ref{fig:profiles_NaD_KD} shows the cross sections for the Na $D$ (blue) and K $D$ (red) lines at $T = 2000$ K for different densities $n_{\rm H_2}$. The solid curves correspond to unified line profiles, while the dotted curves represent the Voigt profiles based on the impact theory. At the lowest density, $n_{\rm H_2} = 10^{21} \ \rm cm^{-3}$, the absorption is still dominated by the line center (i.e. small $\Delta \omega$), and there is reasonable agreement between the two approaches, although the wings are already slightly underestimated by the impact theory. As the density increases, short-range interactions, which produce large frequency shifts $\Delta \omega$, become more frequent, and the line wings grow in strength. Since the impact theory fails to accurately capture short-range interactions, requiring an accurate description of $\Delta V$, it increasingly underestimates the strength of the line wings at higher density.   

Starting from the panel corresponding to $n_{\rm H_2} = 8 \times 10^{21} \ \rm cm^{-3}$ in Fig. \ref{fig:profiles_NaD_KD}, another important property predicted by unified theory becomes apparent. At sufficiently high densities, lines can experience notable shifts away from their unperturbed centers ($\Delta \omega = 0$). In the case of the Na-H$_2$ and K-H$_2$ profiles, the shifts can become large enough for two clear bumps to appear. The red-shifted bumps are associated with the $D1$ lines, while the blue-shifted bumps correspond to the $D2$ lines. The shift in the $D1$ line is caused by pressure broadening increasingly favoring absorption at large frequency shifts $\Delta \omega$. Since the $D1$ lines exhibit only a red wing (i.e. $\Delta \omega < 0$), the absorption is progressively shifted toward larger wavelengths, resulting in a systematic redshift with increasing $n_{\rm H_2}$. At the highest density considered, $n_{\rm H_2} = 5 \times 10^{22} \ \rm cm^{-3}$, the $D1$ lines exhibit significant absorption even beyond $10 \ \rm \mu m$ due to this extreme redshift. 
The effect of line shifts cannot be accurately modeled by the impact theory when using the common Van der Waals potential to construct a Lorentz profile, especially at high densities. Although more accurate interaction potentials can allow more realistic line shifts to be included within the impact theory, as demonstrated by \cite{deRegt_2025} for the $4p-5s$ lines of K broadened by He and H$_2$, the resulting profile remains essentially a shifted Lorentzian. The impact theory is therefore unable to reproduce the strongly asymmetric, non-Lorentzian profiles that emerge at high densities in unified theory.

The shift in the $D2$ lines is closely linked to the behavior of their blue satellite bands. At densities of $n_{\rm H_2} \sim 10^{22} \ \rm cm^{-3}$, the blue wing is dominated by absorption in the first line satellite located near $\Delta \omega = \Delta V_{\rm ext}$, causing the line core and satellite to blend. As density increases further, the probability of collisions involving two perturbers grows, eventually exceeding that of single-perturber interactions \citep{allard1978, royer1978}. As a result, the amplitude of the first satellite decreases, while the second satellite at $\Delta \omega = 2\Delta V_{\rm ext}$ becomes more prominent and eventually dominates the blue wing. This process continues with increasing density, as higher-order collisions become more likely. At sufficiently high densities, the third satellite at $\Delta \omega = 3\Delta V_{\rm ext}$ can dominate the absorption. The increase in the probability of multi-perturber collisions has been theoretically demonstrated by \cite{royer1971} and is illustrated in Fig. 27 of \cite{allard1982}. The resulting variations in satellite amplitudes can be well explained using simplified models such as square-well potentials \citep{allard1978, Allard_1980, Allard_1988}, and were first identified experimentally in alkali spectra by \citet{exton1978} and \citet{Kielkopf_1979}. It is this progressive activation of higher-order satellite lines in the blue wings with increasing density that lead to an overall blue shift of the $D2$ line profiles.

Figure \ref{fig:FWHM_Shift_Nad_KD} summarizes the properties of the line profiles at different temperatures. The left panels show the full width at half maximum (FWHM) obtained using unified line theory (solid and dashed curves) and the predictions of the impact theory (dotted curves). For $n_{\rm H_2} \lesssim  5 \times 10^{21} \ \rm cm^{-3}$, there is reasonable agreement between the two approaches, and the line core can still be described by a Lorentzian profile. However, at higher densities, the impact theory breaks down and the line cores can no longer be approximated by Lorentzian profiles. As a result, the impact theory increasingly underestimates the FWHM with increasing $n_{\rm H_2}$, with the discrepancy becoming more pronounced at higher temperatures. The right panels show the shift of the spectral lines as a function of $n_{\rm H_2}$. The shift is defined as the distance between the unperturbed line center ($\Delta \omega$ = 0) and the peak amplitude of the line in wavenumber space. Around $n_{\rm H_2} \sim 7 \times 10^{21} \ \rm cm^{-3}$, the shift becomes non-negligible across all temperatures for each line, and increases continuously in magnitude with density. As explained above, the $D1$ lines exhibit a negative shift due to their strong red wings, while the $D2$ lines show a positive shift as a result of their blue satellite bands. 

Our computed profiles highlight the complexity of the Na $D$ and K $D$ lines at high densities. Using unified theory is essential under these conditions, as simpler approaches such as the impact theory can significantly underestimate the line widths and fail to reproduce the density-dependent line shifts.

\subsection{Rosseland mean opacities: Unified theory versus impact theory}
In the previous section, we illustrated the significant differences in the Na $D$ and K $D$ line profiles at high perturber densities $n_{\rm H_2}$ when using unified line theory compared to the impact theory. These differences directly affect high-pressure opacity calculations, in particular $\kappa_{\rm R}$, which can be very sensitive to the adopted line profiles. In this section, we investigate how the SM26 mean opacities, denoted as $\kappa_{\rm R}^{\rm imp}$, change when using the revised Na $D$ and K $D$ profiles. The updated opacities are denoted as $\kappa_{\rm R}^{\rm unf}$.

\begin{figure}[!h]
    \centering
    \includegraphics[width=1\linewidth]{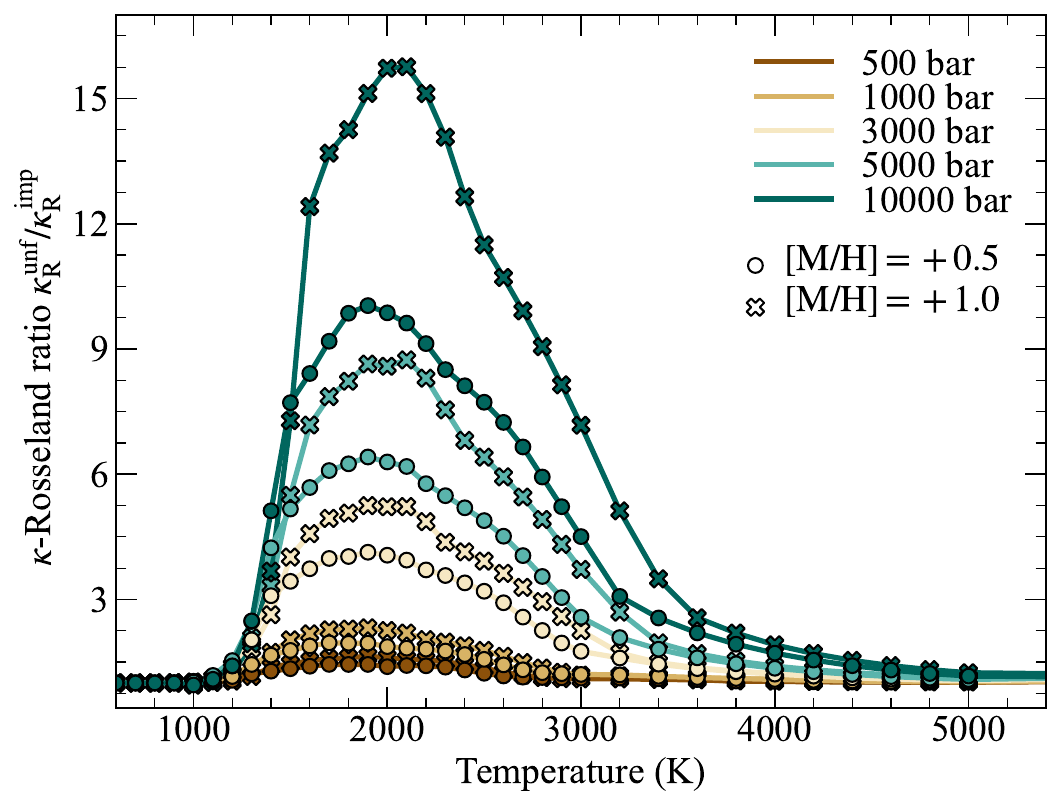}
    \caption{Ratio of the Rosseland mean opacity computed using Na $D$ and K $D$ line profiles from unified line theory, $\kappa_{\rm R}^{\rm unf}$, to that obtained in \cite{Siebenaler_2026}, $\kappa_{\rm R}^{\rm imp}$. Each curve corresponds to a constant pressure. Dots indicate tables evaluated at a metallicity of $\rm [M/H]=+0.5$ ($\sim 3 \times  \rm solar$), while crosses indicate $\rm [M/H]=+1.0$ ($10 \times \rm solar$).}
    \label{fig:ratio_unf_vs_imp}
\end{figure}

In Fig. \ref{fig:ratio_unf_vs_imp}, we show the ratio between $\kappa_{\rm R}^{\rm unf}$ and $\kappa_{\rm R}^{\rm imp}$. In general, mean opacities increase between $\sim 1000 - 4000 \ \rm K$ when using the updated line profiles. This enhancement arises from the stronger and more extended wings of the Na $D$ and K $D$ lines, which efficiently fill otherwise low-opacity regions in the $\sim 0.3 - 2 \ \rm \mu m$ wavelength range. Since the line wings become increasingly important at higher pressures, and the Na and K lines shift further apart, the increase in $\kappa_{\rm R}^{\rm unf}$ relative to $\kappa_{\rm R}^{\rm imp}$ becomes more pronounced with pressure. At pressures of $\sim 10^3 \ \rm bar$, $\kappa_{\rm R}^{\rm unf}$ is larger by a factor of 2, while at $\sim 10^4 \ \rm bar$, it can exceed $\kappa_{\rm R}^{\rm imp}$ by up to an order of magnitude for a metallicity of $\rm [M/H] = +0.5$, with even larger differences at higher metallicities. We note that if pressure further increases the impact of the revised line profiles on mean opacities diminishes since collision induced absorption (CIA) and free-free absorption, and bound-free absorption become the dominant opacity sources.

At a given pressure, the difference between $\kappa_{\rm R}^{\rm unf}$ and $\kappa_{\rm R}^{\rm imp}$ is maximal near $\sim 2000 \ \rm K$. In this region, $B_{\nu}$ peaks around $\lambda \sim 2 \ \rm \mu m$, and Na and K remain in the gas phase, making them abundant contributors to the opacity. At the same time, other short-wavelength absorbing species (e.g., free electrons, metal hydrides, and metal oxides) are not abundantly present yet, making Na and K the dominant sources to absorb at short wavelengths in this region. However, as temperature increases, other short-wavelength absorbing species become more abundant, thus reducing the relative contribution of the Na $D$ and K $D$ lines to the opacity. At temperatures $\lesssim 1000 \ \rm K$, Na and K condense out of the gas phase, making their contribution negligible.

Using the updated Na $D$ and K $D$ line profiles results in more opaque giant planets, which can influence their thermal structure and long-term evolution. The potential implications of these updated mean opacities are explored and discussed in the next section.

\section{Discussion} \label{sec:Discussion}

\subsection{Radiative-convective boundary in warm and hot Jupiters}

\begin{figure}[!h]
    \centering
    \includegraphics[width=1\linewidth]{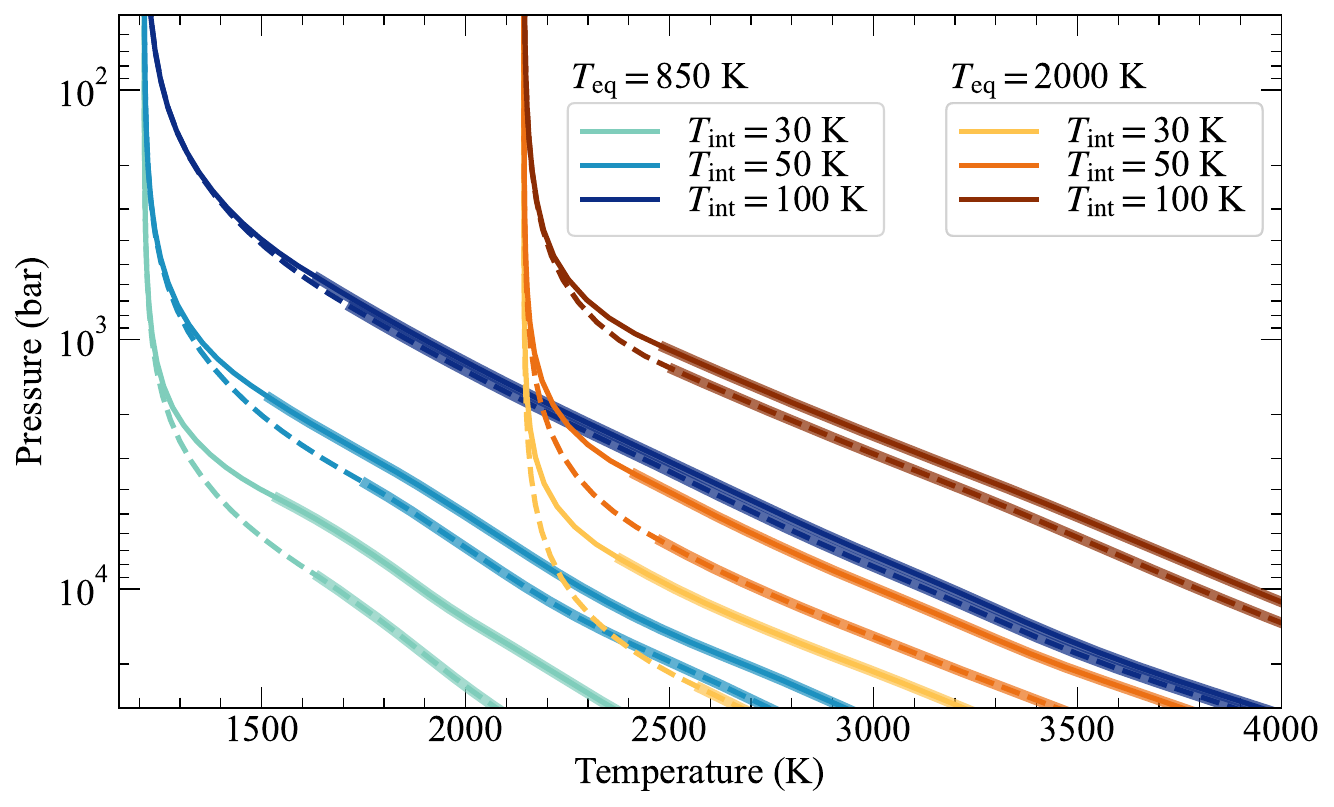}
    \caption{Temperature-pressure profiles of solar atmospheric metallicity models with different $T_{\rm eq}$ and $T_{\rm int}$. The solid curves are obtained using $\kappa_{\rm R}$ based on the revised Na $D$ and K $D$ line profiles from unified theory, while the dashed curves use $\kappa_{\rm R}$ tables where they are modeled as Voigt profiles. The shaded regions indicate where the planet is convective.}
    \label{fig:HJ_profiles}
\end{figure}

\begin{figure}[!h]
    \centering
    \includegraphics[width=1\linewidth]{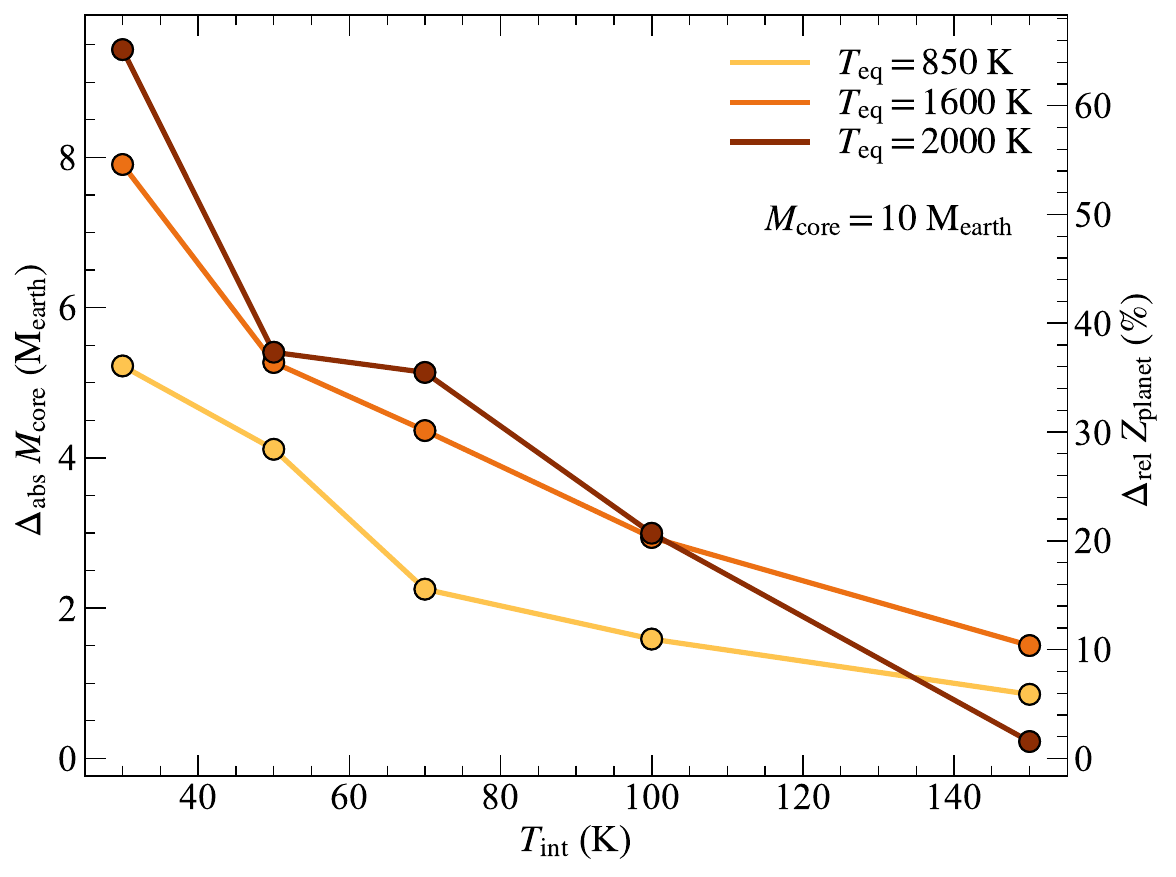}
    \caption{Absolute change in the core mass $M_{\rm core}$ and relative change in the bulk metallicity $Z_{\rm planet}$ as a function of the intrinsic temperature $T_{\rm int}$ when using the revised opacity tables incorporating the Na $D$ and K $D$ profiles from unified theory. We assume an initial core mass $M_{\rm core}=10 \ \rm M_{\rm earth}$. For each model, we assume a compact core and a homogeneous envelope with solar composition. Each curve applies to a different equilibrium temperature $T_{\rm eq}$. }
    \label{fig:Z_change}
\end{figure}

We now examine how the updated Na $D$ and K $D$ line profiles can modify the radiative–convective boundary (RCB) in warm and hot Jupiters. For colder planets, such as Jupiter or Saturn analogs, the RCB is not affected by these updates, as it occurs at temperatures where alkalis are no longer in the gas phase. For planets with sufficiently high equilibrium temperatures ($T_{\rm eq}\gtrsim 500 \ \rm K$), however, the RCB can be located at temperatures $T > 1000 \ \rm K$, where the revised Na $D$ and K $D$ line profiles can significantly affect the opacity. In addition, planets with a low intrinsic heat flux (i.e. a low intrinsic temperature $T_{\rm int}$) can remain radiative deep into the atmosphere, well beyond 100 bar \citep{Guillot_2002, Sudarsky_2003}, where the updated opacities become increasingly important. Hence, warm and hot Jupiters with sufficiently low $T_{\rm int}$ are the primary candidates for shifts in the location of the RCB.

Figure \ref{fig:HJ_profiles} shows temperature profiles of planets with different $T_{\rm eq}$ and varying $T_{\rm int}$. They correspond to static interior models of Jupiter mass planets with a compact core mass of $10 \ \rm M_{\rm Earth}$ produced using \texttt{CEPAM} \citep{Guillot_1995}. We used the \cite{Parmentier_2014} atmosphere model as an atmosphere-interior boundary condition at an optical depth $\tau = 10$, below which the radiative temperature gradient was computed using $\kappa_{\rm R}$. For each model we assume a solar composition. The solid curves are obtained using $\kappa_{\rm R}^{\rm unf}$ tables that incorporate the updated line profiles from unified theory, while the dashed curves use $\kappa_{\rm R}^{\rm imp}$ tables in which the Na $D$ and K $D$ lines are modeled as Voigt profiles when $n_{\rm H_2} \geq 10^{21}\ \rm cm^{-3}$. The shaded regions indicate where the planet is adiabatic, which is modeled using the \cite{Chabrier_2019} H–He equation of state with \cite{Howard_2023} nonideal mixing. 

As anticipated, using the revised opacities $\kappa_{\rm R}^{\rm unf}$ leads to more opaque planets, increasing the radiative gradient and shifting the RCB to lower pressures. At the pressure corresponding to the old RCB, the temperature can increase by up to $20\%$ when using the revised opacities and very low intrinsic temperatures ($T_{\rm int} = 30\ \rm K$). It should be expected that these temperature differences propagate to the center of the planet, since the adiabatic gradients in both models are similar at deeper levels. For high $T_{\rm int}$, the radiative temperature gradient increases and planets will not remain radiative as deep into their interior. This reduces the impact of the revised opacities at higher $T_{\rm int}$. We find that once $T_{\rm int} = 200\ \rm K$ is reached, the temperature profiles obtained with both $\kappa_{\rm R}$ tables become nearly identical for all $T_{\rm eq}$.

In general, planets become warmer as a result of these shifts in the RCB, which will affect their internal structure. While the radius evolution is expected to be largely unaffected, the location of key phase transitions, such as the transition to metallic hydrogen \citep{Sano_2011, Loubeyre_2012}, as well as the inferred bulk metallicities $Z_{\rm planet}$, can change due to the warmer interior. To quantify the effect on inferred bulk metallicities, we determine the core mass $M_{\rm core}$ required to reproduce the same planetary radius in our static interior models using both opacity tables. Since the revised opacity table $\kappa_{\rm R}^{\rm unf}$ produces warmer and therefore puffier interiors, matching the radii obtained with $\kappa_{\rm R}^{\rm imp}$ requires larger core masses and consequently higher bulk metallicities $Z_{\rm planet}$. Figure \ref{fig:Z_change} shows the resulting increase in core mass and bulk metallicity for planets with different $T_{\rm eq}$ and $T_{\rm int}$, assuming an initial core mass of $M_{\rm core}=10 \ \rm M_{\rm earth}$. For planets with $T_{\rm int}=30 \ \rm K$, we find that the inferred core mass can almost double, increasing by up to $\sim 9 \ \rm M_{\rm earth}$ at $T_{\rm eq}=2000 \ \rm K$. This corresponds to a relative increase in bulk metallicity of up to $\sim 66 \ \%$. As $T_{\rm int}$ increases, the impact of the revised opacities on the RCB is reduced, and consequently the change in $Z_{\rm planet}$ as well. We also note that, for denser planets with larger core masses, the relative change in the bulk metallicity will be reduced. This is because the absolute increase in core mass is relatively insensitive to the choice of initial core mass in our models.

An increase in core mass of $9 \ \rm M_{\rm earth}$ is significant, especially in the context of giant planet formation where a core mass of $10 \ \rm M_{\rm earth}$ is classically associated with the onset of runaway gas accretion \citep{Mizuno_1978, Mizuno_1980}. In this regime, opacity-driven shifts of several $M_{\rm earth}$ can have important implications for when runaway gas accretion begins, how efficiently gas is accumulated before disk dispersal, and how the final heavy-element budget is distributed between the core and the envelope. However, the interior models of \cite{vanDijk_2025} indicate that current observational uncertainties remain too large to robustly distinguish opacity-driven shifts in core mass of the magnitude predicted in this study. Nevertheless, the predicted shifts can be large enough to be physically meaningful for formation and evolution models, and may influence how inferred core masses are interpreted in the context of giant planet evolution.

As discussed above, the magnitude of the opacity-driven shifts strongly depends on $T_{\rm int}$, but the appropriate values for irradiated giant exoplanets, particularly hot Jupiters, remain debated. For the Solar System giants, Cassini-based measurements of internal heat fluxes allow to derive $T_{\rm int} = 107.19 \pm 0.57 \ \rm K$ for Jupiter \citep{Li_2018} and $T_{\rm int} = 84.13 \pm 1.48 \ \rm K$ for Saturn \citep{Wang_2024}. In standard cooling models of irradiated giant planets that do not include ongoing anomalous heating, they can cool toward similarly low $T_{\rm int}$ \citep{Guillot_2002,Fortney_2007}, in which case the radiative region can reach $\sim \ \rm kbar$ levels. In contrast, studies that interpret the inflated radii of hot Jupiters as evidence for ongoing deposition of heat into the interior suggest substantially larger $T_{\rm int}\gtrsim 200 \ \rm K$, which can shift the RCB to much lower pressures \citep{Thorngren_2019, Sarkis_2021}. An important caveat, however, is that alternative mechanisms such as wind-driven downward energy advection can also reproduce inflated radii by modifying the thermal structure throughout the radiative region \citep{Tremblin_2017}. In this scenario, the radiative region can extend to greater depths and the inferred $T_{\rm int}$ can be lower than predicted by 1D deep-heating models.

\subsection{Condition for stable radiative layer in Jupiter}

\begin{figure}[!h]
    \centering
    \includegraphics[width=1\linewidth]{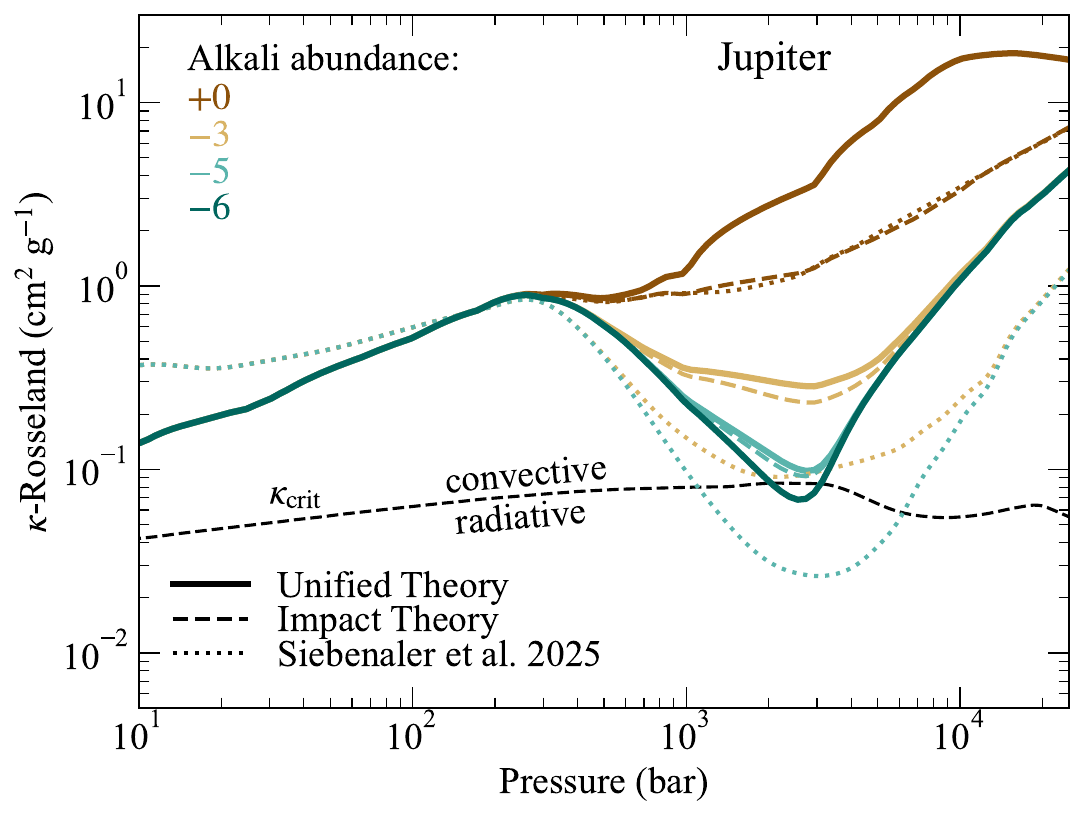}
    \caption{Rosseland mean opacities for present-day Jupiter. Solid lines apply to the new opacities from \cite{Siebenaler_2026} including unified theory treatment. Dotted lines apply to the opacities from \cite{Siebenaler_2025}.}
    \label{fig:Jupiter_Saturn_kr}
\end{figure}

\begin{figure}[!h]
    \centering
    \includegraphics[width=1\linewidth]{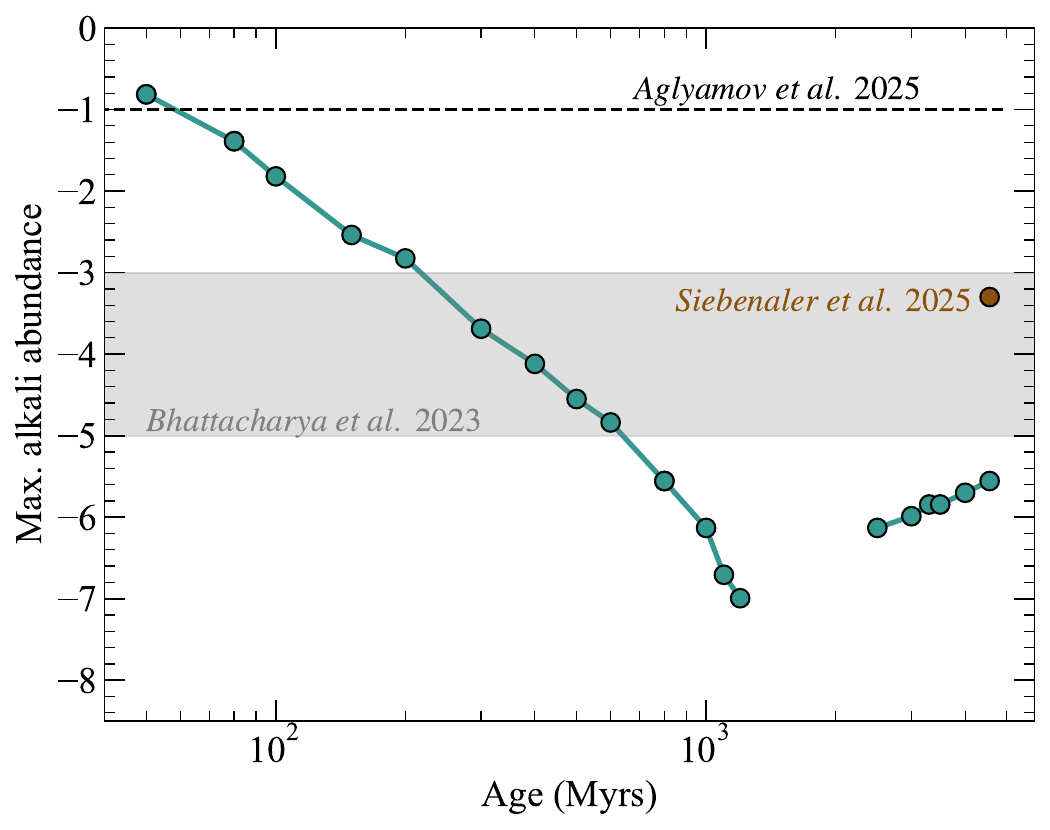}
    \caption{Maximum Na and K abundance relative to solar that allows Jupiter to sustain a stable radiative zone as a function of time. The dark cyan curve indicates the results from this study, while the dark point gives the condition for a stable radiative zone from \cite{Siebenaler_2025}. The black dashed curve and gray shaded region give observational constraints on Jupiter alkali abundance from \cite{Aglyamov_2024} and \cite{Bhattacharya_2023} respectively.}
    \label{fig:Jupiter_evolution}
\end{figure}

It has been previously shown that the Na $D$ and K $D$ lines play a key role in regulating the temperature gradients in Jupiter’s H$_2$-envelope (e.g., \citealp{Guillot_2004}; S25). In S25, we found that an alkali depletion below $10^{-3}$ times solar can lead to the formation of a stable radiative zone located between $\sim 1-7\ \rm kbar$. Here, we revisit this condition using our updated opacity tables and extend the analysis to Jupiter’s evolution. 

In Fig.~\ref{fig:Jupiter_Saturn_kr}, we show gas-only $\kappa_{\rm R}$ computed along a dry adiabatic Jupiter thermal profile for different Na and K abundances. The solid curves correspond to $\kappa_{\rm R}^{\rm unf}$, while the 
dashed curves show $\kappa_{\rm R}^{\rm imp}$. The dotted curves correspond to the results from S25. Using $\kappa_{\rm R}^{\rm unf}$, we find that the alkali abundance required to sustain a stable radiative layer at the present time is reduced to below $10^{-5}$ times solar, significantly lower than the $10^{-3}$ condition reported in S25. This is well below current Juno mission constraints of Jupiter’s alkali content \citep{Bhattacharya_2023, Aglyamov_2024}, making the presence of such a stable region due to opacity reduction highly unlikely. In addition, the resulting radiative zone is narrower, extending only over $\sim 2-3\ \rm kbar$. This is also narrower than the radiative zone reported by \cite{Muller_2024}, which uses an ad hoc scaling of the opacities to permit sub-adiabatic layers. 

Importantly, neither the stronger alkali depletion requirement nor the reduced extent of the potential radiative region is driven by the updated Na $D$ and K $D$ line profiles. Comparing $\kappa_{\rm R}^{\rm unf}$ (solid curves) and $\kappa_{\rm R}^{\rm imp}$  (dashed curves) shows negligible differences for alkali abundances $\lesssim 10^{-3}$ times solar. Significant deviations appear only at higher alkali abundances ($\gtrsim 10^{-3}$ times solar). Instead, the differences in our revised opacities to S25 at low alkali abundances are primarily driven by updates to the 
H$_2$–H$_2$ CIA data. S25 uses CIA data from \cite{Abel_2012}, which extends down to $1\ \mu\rm m$, whereas SM26 combines data from \cite{Borysow_2001}, \cite{Borysow_2002}, and \cite{Abel_2012}, extending the wavelength coverage down to $0.5\ \mu\rm m$. This additional short-wavelength contribution increases $\kappa_{\rm R}$ at temperatures $\gtrsim 1200\ \rm K$ under conditions of alkali depletion. 

Another difference in the SM26 opacity dataset is the conservation of the integrated line flux when a line-wing cutoff is applied, following the approach of \cite{Sharp_2007}. This was not enforced in S25, leading to systematically weaker molecular and atomic opacities at high pressures. At lower pressures, $\lesssim 100 \ \rm bar$, additional differences between the SM26 and S25 opacity data arise from the adopted chemistry networks. SM26 relies on \texttt{GGchem}, whereas S25 uses \texttt{FastChem Cond} \citep{Kitzmann_2023}. Since \texttt{FastChem Cond} includes fewer condensates, PH$_3$ is not removed from the gas phase at low temperatures, which explains the higher gas opacity in S25 in this regime.

Figure \ref{fig:Jupiter_evolution} shows the alkali depletion required for the formation of a stable layer as a function of time. We used \texttt{CEPAM} to generate dry adiabatic temperature profiles of a Jupiter-like planet as function of time and computed $\kappa_{\rm R}$ along these profiles for different alkali abundances. The planet is assumed to have a $10 \rm \ M_{\oplus}$ core composed of $50\%$ rock and $50\%$ ice, and is surrounded by a homogeneous H–He envelope of protosolar composition. For the atmosphere-interior boundary condition, we use the tables from \cite{Fortney_2011}. 

We find that a stable layer below the RCB only forms after $\sim 50 \ \rm Myrs$ into Jupiter's evolution. At earlier times, the region where alkali begin to form remains above the RCB. Over most of Jupiter’s evolution, a stable layer could in principle exist if there is a sufficient alkali depletion. At early times, however, a stable layer can form at less extreme alkali depletion. The planet is hotter, which shifts the potential stable layer to lower pressures. Under these conditions, the Na $D$, K $D$, and H$_2$–H$_2$ CIA opacities are reduced, which lowers the radiative gradient. As a result, a higher alkali abundance is required to maintain convection. This reduction in opacity outweighs the effect of the higher internal temperature $T_{\rm int}$, which would otherwise increase the radiative gradient. As the planet evolves and cools, the potential stable layer shifts deeper into the interior, and the alkali abundance required to sustain convection decreases. Between $\sim 1200 - 2500\ \rm Myr$, the atmosphere becomes sufficiently opaque that a stable region cannot form, even if alkali are fully removed. At later times, however, the continued decrease in $T_{\rm int}$ again allows for the formation of a stable layer under sufficiently depleted conditions. 

Despite Jupiter's early conditions favoring the formation of stable layers, given current estimates of its alkali abundances, it is highly unlikely that it hosted a stable radiative region at any stage of its evolution through this opacity reduction mechanism. The only possible window would be around $50 - 60\ \rm Myr$, and even this may disappear once condensate opacities are considered. 
However, this analysis does not fully rule out the presence of stable layers in Jupiter’s hydrogen envelope. At deep levels, Jupiter could be sufficiently conductive to inhibit convection and become stable. In such cases, an accurate description of the Na $D$ and K $D$ opacities at high pressures remains essential. Alternatively, Jupiter may host shallower superadiabatic stable layers near the water-cloud condensation level \citep{Li_2024}.

Although our analysis suggests that Jupiter's envelope cannot host a stable layer as result of an alkali depletion, certain exoplanets may still host non-convective regions due to this mechanism. \cite{Muller_2026} show that warm Jupiters with equilibrium temperatures of $200-800$ K can develop radiative zones using their ad hoc opacity scaling approach, consistent with our finding that warmer conditions favor stable layers. Based on our results, we suggest that the most favorable environments for such layers are warm, old giant planets (i.e. with low $T_{\rm int}$). However, we stress that determining their extent and whether they can form at all requires detailed opacity calculations similar to those presented here.

\section{Conclusions} \label{sec:Conclusion}

We have computed new Na $D$ and K $D$ absorption cross sections using unified line theory, extending to  H$_2$ perturber densities of $n_{\rm H_2} = 5 \times 10^{22} \ \rm cm^{-3}$, thereby enabling improved opacity calculations at high pressures. We have incorporated these calculations into mean opacity tables relevant for giant planet interiors. 

We find that for $n_{\rm H_2}>10^{21} \ \rm cm^{-3}$, the Na $D$ and K $D$ lines develop significantly stronger and more extended wings compared to commonly used Voigt profiles with large line-wing cutoffs. In addition, the line centers shift at high densities $n_{\rm H_2} \gtrsim 8 \times 10^{21} \ \rm cm^{-3}$, with the $D1$ lines exhibiting a redshift  due to their strong red wing, and the $D2$ lines showing a blueshift associated with their blue satellite bands. As a result, $\kappa_{\rm R}$ increases over temperatures of $\sim 1000 - 4000 \ \rm K$, and can exceed previous calculations by a factor of 2  at pressures of $10^{3} \ \rm bar$ and by more than one order of magnitude at pressures of $10^{4} \ \rm bar$.

We demonstrated that these opacity changes can modify the thermal structure of warm and hot giant planets ($T_{\rm eq} \gtrsim 500 \ \rm K$) with sufficiently low intrinsic temperatures ($T_{\rm int} < 200 \ \rm K$). The enhanced opacities increase the radiative temperature gradient, shifting the RCB to lower pressures and leading to a warmer interior adiabat. As a result, we estimate that inferred core masses can increase by upto $9 \ \rm M_{\rm earth}$, which is large enough to be physically meaningful for formation and evolution models.

For Jupiter, we revisited the condition under which an alkali depletion driven stable radiative layer can form. Using our updated opacity tables, we find that an elemental abundance of K and Na below $\sim 10^{-5}$ times solar is required to sustain such a  radiative zone at the present time, substantially lower than previously estimated. This is well below observational constraints of Jupiter's alkali abundance, making the presence of a stable radiative region at kilobar pressures due to an opacity reduction highly unlikely. Importantly, we show that this revised condition is primarily driven by updates to H$_2$–H$_2$ CIA, rather than by changes in the Na $D$ and K $D$ line profiles. We also explored the possibility of such a layer throughout Jupiter’s evolution. Although we find that a stable layer is more easily formed at early times due to the planet's higher effective temperature, it is unlikely that Jupiter hosted a persistent stable radiative region as a result of an opacity reduction over most of its evolution. However, while Solar System giant planets are unlikely to host these structures, warm giant planets with low intrinsic heat fluxes remain promising environments for stable layers, for which detailed alkali line profiles as presented in this work become essential.  

In general, this work highlights the importance of accurately modeling the Na $D$ and K $D$ line profiles at high densities for determining the thermal structure of giant planets. While our analysis focused on these four lines, significant uncertainties remain in both atomic and molecular lines in high-density environments. The common practice of adopting Voigt profiles with ad hoc line-wing cutoffs at all pressures is not well justified and can lead to incorrect opacity estimates, even at smaller pressures considered in this study. Although computing autocorrelation functions for all transitions is not feasible, improving our understanding of appropriate line-wing treatments as a function of pressure would substantially reduce these uncertainties. However, at high densities the impact of other atomic and molecular lines on $\kappa_{\rm R}$ is expected to be subdominant compared to the Na $D$ and K $D$ lines, and CIA and free electron opacities. We therefore conclude that the present work addresses the most important known absorption lines for high-pressure opacity calculations in giant planets.

\section*{Data availability}

The mean opacities and the Na $D$ and K $D$ cross sections used in this study are available at \url{https://doi.org/10.5281/zenodo.20794489}. The autocorrelation functions used throughout this studies will be provided upon reasonable request.

\begin{acknowledgements}
      We thank the referee for valuable comments which helped improve the manuscript. This project has received funding from the European Research Council (ERC) under the European Union’s Horizon 2020 research and innovation programme (grant agreement no. 101088557, N-GINE). This publication is part of the project ENW.GO.001.001 of the research programme “Use of space infrastructure for Earth observation and planetary research (GO), 2022-1” which is (partly) financed by the Dutch Research Council (NWO). NFA  benefited from support from CNES as part of the Ariel space mission. We thank Tristan Guillot for insightful discussions and encouragement of this work.
\end{acknowledgements}

%
%

\bibliographystyle{aa} 
\bibliography{bibliography.bib} 

\clearpage
\onecolumn
\begin{appendix}

\section{Metal abundances $\rm [M/H]$}

\begin{figure}[!h]
    \centering
    \includegraphics[width=0.5\linewidth]{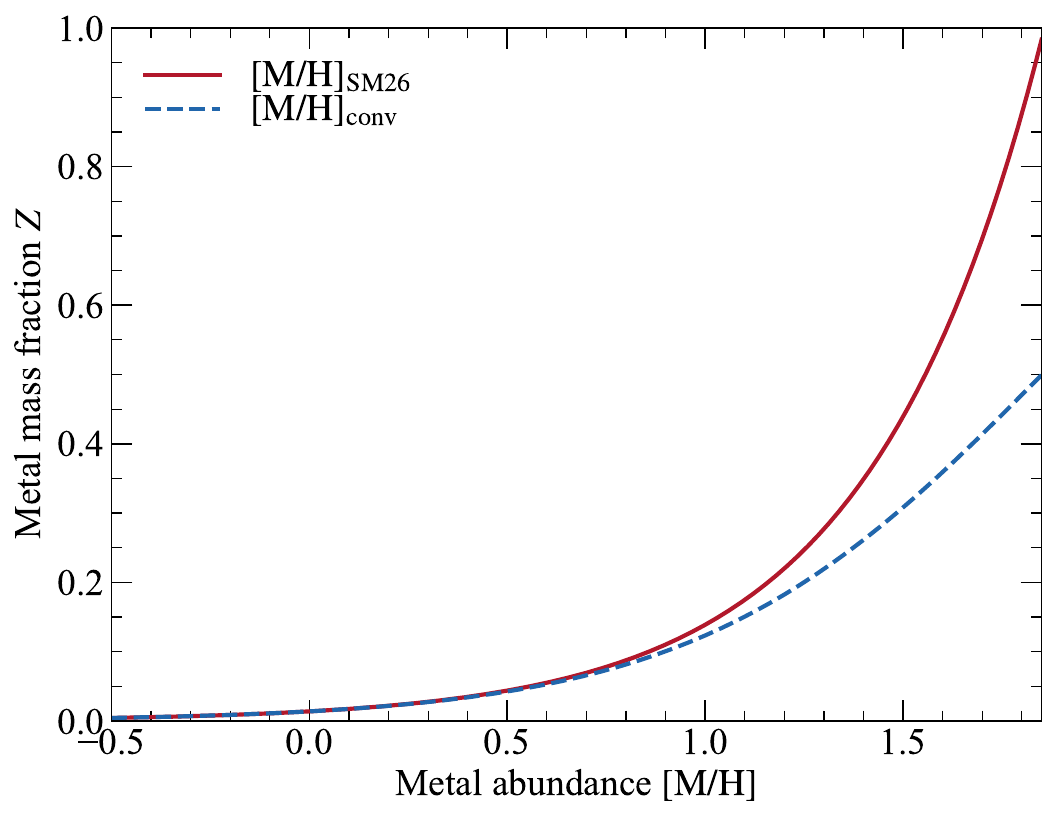}
    \caption{Metal mass fraction $Z$ as a function of the metal abundance $\rm [M/H]$. In red we show the $\rm [M/H]$ definition that was adopted in SM26, and in blue we show the conventional definition.}
    \label{fig:Z_vs_[M_H]}
\end{figure}

In SM26, the chemistry calculations are based on an alternative approach in which the hydrogen (H) and helium (He) abundances, $\rm \log (\epsilon_H)$ and $\rm \log (\epsilon_{He})$, are rescaled such that the H to He mass ratio is fixed to $X/Y = 0.326$. In addition, the total mass is conserved, i.e. $M_{\rm H} + M_{\rm He} + M_{\rm Z} = M_{\rm H,\odot} + M_{\rm He,\odot} + M_{\rm Z,\odot}$. In the conventional approach, the latter is not enforced, and instead, $\rm log \ \epsilon_H$ and $\rm log \ \epsilon_{He}$ are fixed to their solar values, and the total mass becomes $M_{\rm H, \odot} + M_{\rm He, \odot} + M_{\rm Z}$. Depending on the adopted approach, this leads to different interpretations of the metal abundance parameter $\rm [M/H]$. Figure \ref{fig:Z_vs_[M_H]} illustrates how these definitions result in different relationships between $\rm [M/H]$ and the metal mass fraction $Z$. The difference is negligible at $\rm [M/H] \lesssim 1$, but becomes significant at high metallicities. Table \ref{table:abundances} summarizes the abundances $\rm [M/H]_{\rm SM26}$ used in SM26 and their corresponding $Z$. For comparison, we also provide the conventional definition $\rm [M/H]_{\rm conv}$ that reproduces the same $Z$ values. To ensure consistency, we adopt the $\rm [M/H]_{\rm conv}$ definition from now on when reporting our mean opacity tables and provide the corresponding $Z$. These updates are included in the Zenodo repository.

\begin{table}[h]
\small
\centering
 \caption{Metal abundances $\rm [M/H]$ and the corresponding metal mass fraction $Z$.}
 \begin{tabular}{c @{\hskip 5cm }  c  @{\hskip 5cm }  c} 
 \hline
 \hline
$\rm [M/H]_{\rm SM26}$ & $\rm [M/H]_{\rm conv}$  & $Z$ \\ [0.01ex] 
  \noalign{\smallskip}
    \hline
    \noalign{\smallskip}
    -0.5 & - & 0.0044 \\
    -0.3 & - & 0.0070 \\
    +0.0 & - & 0.0139 \\
    +0.3 & - & 0.0277 \\
    +0.5 & +0.51 & 0.0439 \\
    +0.7 & +0.72 & 0.0696 \\
    +1.0 & +1.05 & 0.1389 \\
    +1.5 & +1.75 & 0.4391 \\
    +1.7 & +2.21 & 0.6960 \\
         
  \hline

\end{tabular}
\tablefoot{$\rm [M/H]_{\rm SM26}$ denotes the definition adopted in SM26, and $\rm [M/H]_{\rm conv}$ is the conventional definition.}
 
 \label{table:abundances}

\end{table}

\newpage
\section{Cross section tables} \label{Appendix:xsec}

\begin{table}[h]
\small
\centering
 \caption{Molecular opacities used in this work.}
 \begin{tabular}[!b]{c @{\hskip 0.8cm } c @{\hskip 0.8cm } c @{\hskip 1.5cm } c @{\hskip 1.3cm } c @{\hskip 0.3cm } c} 
 \hline
 \hline
Molecule  & $T_{\rm max}$ (K) & Wavelength ($\mu \rm m$) & Line List Name &  References \\ [0.01ex] 
  \noalign{\smallskip}
    \hline
    \noalign{\smallskip}
    AlH & $5000$ & $0.407 - 500$ & AloHa &\textrm{\cite{Sergei_2023}}  \\
    CaH & $5000$ & $0.335 - 500$ & XAB &\textrm{\cite{Owens_2022a}}  \\
    CaO & $6000$ & $0.400 - 500$ & VBATHY &\textrm{\cite{Yurchenko_2016}}  \\
    CaOH & $5000$ & $0.278 - 500$ & OYT6 &\textrm{\cite{Owens_2022}}  \\
    CH & $6000$ & $0.255 - 200$ & MoLLIST &\textrm{\cite{Masseron_2014, Bernath_2020}}  \\
    CH$_4$ & $5000$ & $0.833 - 500$ & MM & \textrm{\cite{Yurchenko_2024}}  \\
    CO & $6000$ & $0.455 - 500$ & Li2015 &\textrm{\cite{Li_2015, Somogyi_2021}}  \\
    CO$_2$ & $5000$ & $0.500 - 500$ & UCL-4000 &\textrm{\cite{Yurchenko_2020}}  \\
    CP & $3000$  & $0.661 - 28$ & MoLLIST &[1] \\
    CrH & $3000$  & $0.667 - 1.615$ & MoLLIST & [2]  \\
    FeH & $6000$  & $0.667 - 50$ & MoLLIST &\textrm{\cite{Dulick_2003, Bernath_2020}} \\
    H$_2$ & $6000$ & $0.278 - 200$ & RACPPK &\textrm{\cite{Roueff_2019}}\\
    H$_2$O & $6000$ & $0.243 - 500$ & POKAZATEL &\textrm{\cite{Polyansky_2018}}\\
     H$_2$S & $3000$ & $0.286 - 500$ & AYT2 &\textrm{\cite{Azzam_2016, Chubb_2018}}\\
    HCl & $5000$ & $0.494 - 500$ & HITRAN-HCl &\textrm{\cite{Gordon_2017}}\\
    HCN & $4000$ & $0.569 - 500$ & Harris &\textrm{\cite{Harris_2006, Barber_2013}}\\
    HF & $5000$ & $0.31 - 500$ & Coxon-Hajig & [3]\\
    LiOH & $5000$ & $1 - 500$ & OYT7 &\textrm{\cite{Owens_2024}}\\
    MgH & $5000$ & $0.338 - 500$ & XAB &\textrm{\cite{Owens_2022a}}\\
    MgO & $5000$ & $0.270 - 500$ & LiTY &\textrm{\cite{Li_2019}}\\
    N$_2$ & $6000$ & $0.179 - 500$ & WCCRMT & [4]\\
    NaCl & $3000$  & $4.069 - 500$ & Barton &\textrm{\cite{Barton_2014}}\\
    NaH & $6000$ & $0.311 - 500$ & Rivlin &\textrm{\cite{Rivlin_2015, Chubb_2020}}\\
    NH$_3$ & $2000$ & $0.500 - 500$ & CoYuTe &\textrm{\cite{Derzi_2015, Coles_2019}}\\
    PH$_3$ & $3000$ & $1 - 500$ & SAITY &\textrm{\cite{Silva_2014}}\\
    PN & $5000$ & $0.121 - 500$ & PaiN &\textrm{\cite{Semenov_2024}}\\
    PS & $5000$ & $0.270 - 500$ & POPS &\textrm{\cite{Prajapat_2017}}\\
    SiH & $5000$ & $0.313 - 500$ & SiGHTLY &\textrm{\cite{Yurchenko_2017}}\\
    SiH$_4$ & $2000$ & $2 - 500$ & OY2T &\textrm{\cite{Owens_2017}}\\
    SiO & $6000$  & $0.139 - 500$ & SiOUVenIR &\textrm{\cite{Yurchenko_2021}}\\
    SO & $5000$ & $0.222 - 500$ & SOLIS &\textrm{\cite{Brady_2023}}\\
    TiH & $4800$ & $0.417 - 2.156$ & MoLLIST &\textrm{\cite{Burrows_2005, Bernath_2020}}\\
    TiO & $6000$ & $0.333 - 500$ & Toto &\textrm{\cite{McKemmish_2019}}\\
    VO & $5400$ & $0.222 - 500$ & HyVO &\textrm{\cite{Bowesman_2024}}\\
        
  \hline
\end{tabular}
 \label{table:molecule_opacity}

\tablebib{[1]: \textrm{\cite{Ram_2014, Bernath_2020, Qin_2021}} [2]: \textrm{\cite{Burrows_2002, Chubb_2018, Bernath_2020}} [3]: \textrm{\cite{Li_2015, Coxon_2015, Somogyi_2021}} [4]: \textrm{\cite{Shemansky_1969, Western_2017, Western_2018, Jans_2024}}}
 
\end{table}

\begin{table}[h]
\small
\centering
 \caption{Atomic opacities used in this work.}
 \begin{tabular}{c @{\hskip 1.8cm }  c @{\hskip 2.2cm } c} 
 \hline
 \hline
Atom & Wavelength ($\mu$m) & References \\ [0.01ex] 
\hline
\noalign{\smallskip}
   Ca & $0.138 - 500$ & NIST$^{[1]}$; VALD$^{[2]}$
   \\
   Cr & $0.151 - 500$ & NIST; VALD
   \\
   Fe & $0.1 - 65.591$ & \cite{Kurucz_2018}
   \\
   K & $0.299 - 100$ & NIST; VALD; \cite{Allard_2016, Allard_2025}; This study 
   \\
   Li & $0.234-65.119$ & NIST; VALD
   \\
   Mg & $0.162-1.209$ & NIST; VALD
   \\
   Mn & $0.299-28.482$ & NIST; VALD
   \\
   Na & $0.243 - 100$ & NIST; VALD; \cite{Allard_2019, Allard_2025}; This study
   \\
   Ni & $0.170 - 500$ & NIST; VALD 
   \\
   Ti & $0.203 - 2.387$ & NIST; VALD
   \\
   V & $0.201 - 500$ & NIST; VALD
   \\
  \hline
\end{tabular}
 \label{table:atomic_opacity}
 
\tablebib{[1]: \textrm{\cite{NIST_2001}} [2]: \textrm{\cite{Ryabchikova_2017}}}

\end{table}

\newpage

\begin{table}[!h] 
\small
\centering
 \caption{Collision-induced absorption used in this work.}
 \begin{tabular}[t]{c @{\hskip 1.25cm }  c @{\hskip 1.25cm } c @{\hskip 0.4cm } c } 
 \hline
 \hline
Species  & Temperature range & Wavelength ($\mu \rm m$) & References \\ [0.01ex] 
  \noalign{\smallskip}
    \hline
    \noalign{\smallskip}
    H$_2$-H$_2$ & $100 - 400$  & $0.5 - 500$ & \cite{Borysow_2002, Fletcher_2018, Orton_2025}  \\
    & $400 - 3000$ &  & \cite{Borysow_2001, Borysow_2002, Abel_2012}
    \\
    & $3000 - 5000$ &  & \cite{Borysow_2001}
    \\
    H$_2$-He & $100 - 200$ & $0.5 - 500$ & \cite{Borysow_1989, Borysow_1989_2, Orton_2025} \\
    & $200 - 6000$ & & \cite{Abel_2011}
    \\
    H$_2$-H & $1000 - 2500$ & $1 - 100$ & \textrm{\cite{Gustafsson_2003}} \\
    H$_2$-CH$_4$ & $100 - 400$ &  $5.139 - 500$ & \textrm{\cite{Borysow_1986}}  \\
    H$_2$-CO$_2$ & $200 - 350$ &  $5 - 500$ & \textrm{\cite{Wordsworth_2017}}  \\
    He-H & $1500 - 6000$ &  $0.9 - 200$ & \textrm{\cite{Gustafsson_2001}}  \\
    He-CH$_4$ & $100 - 350$ &  $10 - 500$ & \textrm{\cite{Taylor_1988}}  \\
    CH$_4$-CH$_4$ & $100 - 400$ &  $10.1 - 500$ & \textrm{\cite{Borysow_1987}}  \\

  \hline
\end{tabular}
 \label{table:CIA_opacity}

\end{table}

\begin{table}[!h]
\small
\centering
 \caption{Free-free and bound-free absorptions considered in this work.}
 \begin{tabular}{c @{\hskip 2cm }  c @{\hskip 2cm } c @{\hskip 0.cm}  } 
 \hline
 \hline
Reaction & Wavelength ($\mu$m) & References \\ [0.01ex] 
  \noalign{\smallskip}
    \hline
    \noalign{\smallskip}
    $\textrm{H}_2 + \textrm{e}^- + \textrm{h}\nu \xrightarrow{} \textrm{H}_2 +  \textrm{e}^-$ & $0.351 - 500$ &  \textrm{\cite{Bell_1980}}  \\
     $\textrm{H} + \textrm{e}^- + \textrm{h}\nu \xrightarrow{} \textrm{H} +  \textrm{e}^-$ & $0.182 - 500$ &\textrm{\cite{John_1988}}  \\
     $\textrm{He} + \textrm{e}^- + \textrm{h}\nu \xrightarrow{} \textrm{He} +  \textrm{e}^-$ & $0.506 - 500$ & \textrm{\cite{John_1994}}  \\
     $\textrm{Li} + \textrm{e}^- + \textrm{h}\nu \xrightarrow{} \textrm{Li} +  \textrm{e}^-$ &$0.5 - 500$ & \textrm{\cite{John_1975}}  \\
     $\textrm{N} + \textrm{e}^- + \textrm{h}\nu \xrightarrow{} \textrm{N} +  \textrm{e}^-$ &  $0.5 - 500$ & \textrm{\cite{John_1975}}  \\
     $\textrm{O} + \textrm{e}^- + \textrm{h}\nu \xrightarrow{} \textrm{O} +  \textrm{e}^-$ &  $0.5 - 500$ & \textrm{\cite{John_1975}}  \\
     $\textrm{Na} + \textrm{e}^- + \textrm{h}\nu \xrightarrow{} \textrm{Na} +  \textrm{e}^-$ & $0.5 - 500$ & \textrm{\cite{John_1975}}  \\
     $\textrm{CO} + \textrm{e}^- + \textrm{h}\nu \xrightarrow{} \textrm{CO} +  \textrm{e}^-$ & $0.1 - 500$ & \textrm{\cite{John_1975}}  \\
     $\textrm{N}_2 + \textrm{e}^- + \textrm{h}\nu \xrightarrow{} \textrm{N}_2 +  \textrm{e}^-$ & $0.1 - 500$ & \textrm{\cite{John_1975}}  \\
     $\textrm{H}_2\textrm{O} + \textrm{e}^- + \textrm{h}\nu \xrightarrow{} \textrm{H}_2\textrm{O} +  \textrm{e}^-$ & $0.1 - 500$ & \textrm{\cite{John_1975}}  \\
     $\textrm{H}^-  + \textrm{h}\nu \xrightarrow{} \textrm{H} +  \textrm{e}^-$ & $0.1 - 1.644$ &  \textrm{\cite{McLaughlin_2017}}  \\
  \hline
\end{tabular}
 \label{table:BF_FF_opacity}

\end{table}

\end{appendix}

\end{document}